# An Ab-initio Study of the Y Decorated 2D Holey Graphyne for Hydrogen Storage Application

*Mukesh Singh[a], Alok Shukla[a], Brahmananda Charkraborty[b,c,*]*,

[a]Department of Physics, Indian Institute of Technology Bombay, Powai, Mumbai 400076, India

[b]High Pressure and Synchrotron Radiation Physics Division, Bhabha Atomic Research Centre, Trombay, Mumbai, India

[c]Homi Bhabha National Institute, Mumbai, India

**Abstract:** Expanding pollution and rapid consumption of natural reservoirs (gas, oil, and coal) led humankind to explore alternative energy fuels like hydrogen fuel. Solid-state hydrogen storage is most desirable because of its usefulness in the onboard vehicle. In this work, we explored the yttrium decorated ultra porous, two-dimensional holey-graphyne for hydrogen storage. Using the first principles DFT simulations, we predict that yttrium doped holey graphyne can adsorb up to seven hydrogen molecules per yttrium atom resulting in a gravimetric hydrogen weight percentage of 9.34, higher than the target of 6.5 wt% set by the US Department of Energy (DoE). The average binding energy per $H_2$ and desorption temperature come out to be -0.34 eV and ~ 438 K, respectively. Yttrium atom is bonded strongly on HGY sheet due to charge transfer from Y 4d orbital to C 2p orbital whereas the adsorption of $H_2$ molecule on Y is due to Kubas-type of interactions involving charge donation from H 1s orbital to Y 3d orbital and back donation with net charge gain by H 1s orbital. Furthermore, sufficient energy barriers for the metal atom diffusion have been found to prevent the clustering of transition metal (yttrium) on HGY sheet. The stability of the system at higher temperatures is analyzed using Ab-initio molecular dynamics (AIMD) method, and the system is found to be stable at room and the highest desorption temperature. Stability of the system at higher temperatures, presence of adequate diffusion energy barrier to prevent metal-metal clustering, high gravimetric wt% of $H_2$ uptake with suitable binding energy, and desorption temperature signifies that Y doped HGY is a promising material to fabricate high capacity hydrogen storage devices.

*Corresponding author: Dr. Brahmananda Chakraborty
E-mail address: brahma@barc.gov.in

# 1. Introduction:

Constantly expanding energy requirements are not sustainable with rapidly depleting conventional fossil fuels. In 2009, Shafiee et al. estimated that traditional fuels like gas, oils, and coals would diminish to zero in ~ 35, 37, and 107 years, respectively, so only coal will remain up to the year 2112 [1]. Use of fossil fuels generates hazardous gases like $CO_2$, CO, etc., putting health and lives at risk and polluting the environment. The effects of hazardous gases and their suitable detecting materials are proposed [2,3]. Alternative energy sources must be explored because of fossil fuel depletion and the adverse impact of polluting gases. Important alternavitives resources include solar energy, wind energy, hydrogen energy, etc., and mechanism for optimizing solar energy with iron based catalysts has reported by Zhenzhen et al [4]. The US Department of Energy (DoE) has systematically listed several alternative energy sources [5] out of which hydrogen appears to be extremely promising because of its high abundance, lightweight, high energy density (143 $MJKg^{-1}$)[6], and zero production of greenhouse gases on usage. However, hydrogen energy poses challenges both in production as well as in storage. Numerous methods such as methan decomposition, electrocatalytic etc have been proposed for hydrogen productions [7]. Here we will be dealing with the second problem, i.e., the storage of hydrogen. DoE has specified several criteria for each method of hydrogen storage [8–14]. But for our purpose, only two of them are most important: (i) the binding energy of adsorbed hydrogen must lie between 0.2-0.7 eV per $H_2$, and (ii) the hydrogen storage weight percentage should be more than 6.5. Historically many different methods and procedures have been discovered to stock hydrogen in various forms [15]. Conventional methods include highly pressurized hydrogen gas and cryogenic liquid $H_2$ gas. But these methods are not efficient and economical, especially for onboard storage, because of two main reasons: first, safety issues, and second, the creation of environments with high-pressure and low-temperature are quite expensive [16,17]. Solid-state hydrogen storage is an alternative to the aforementioned techniques. Historically, many materials have been explored to store hydrogen in the solid-state form, ranging from high-density metal hydrides to lightweight carbon nanomaterials. The high reactivity of hydrogen towards metals and alloys leads to the formation of binary metal hydrides $MH_x$ (M denotes any metal) and intermetallic hydrides $AB_nH_x$ (here A is a rare earth or an alkaline earth metal, and B denotes a transition metal) under normal temperature and pressure conditions [18]. Some examples of hydrides of type $MH_x$ are LiH, $MgH_2$, etc., and those of type $AB_nH_x$ are $Mg_2NiH_4$, $LaNiH_6$, $TiFeH_2$, etc. One of the crucial properties of metal hydrides is high volumetric density, e.g., $LaNiH_6$ achieves volumetric hydrogen density up to 115Kg/$m^3$. $MgH_2$ has been reported to have a gravimetric capacity of 7.7 wt% along with the highest energy density (9 MJ/Kg) [19]. So metal hydride can cater to both high volumetric and gravimetric density of hydrogen storage medium. However, the hydrogen binding in metal hydrides is chemisorption in nature, so desorption of hydrogen starts at a very high temperature in the process of recycling.

Metal organic framework (MOFs)[20] and zeolites are other well-known classes of materials that have been explored for energy as well as hydrogen storage [21]. They are expected to be efficient for hydrogen storage due to their high surface area, micropores and cavities, open metal sites, and chemically tunable structures. However, these properties are favorable at low temperatures, e.g., MOF-5 gives a measure of 4.5 wt% at 78 K and 1 wt% at room temperatures with a pressure of 20 bar [22]. One way to boost the hydrogen storage capability is doping the MOFs with metals. Metal doping increases hydrogen storage up to DoE's target value but only at a lower temperature. Similarly, Zeolite capacity is reported 1.81 wt% at 77 K and 15 bar [23]. As a result, they are also not suitable for operation at room temperature.

Nanomaterials are of great interest due to features of high volumetric density, lightweight, large surface area for adsorption, and chemical stability. Interaction between hydrogen and most of the nanomaterials is dominated by weak van der Waal interaction. Fullerenes, activated carbons, carbon nanotubes (CNT), graphene, graphyne, graphite nanofibers, and other carbon materials can be considered for hydrogen storage because of the above remarkable features [24]. For porous carbon materials and MOFs, gravimetric hydrogen density and surface are linearly correlated. Klechikov et al., in 2015, demonstrated a linear correlation between Brunauer-Emmett-Teller (BET) surface area and hydrogen adsorption wt% for different materials [25]. The hydrogen storage capacity of one of the most explored members of carbon nanomaterial is a carbon nanotube, with its 0.1 wt% uptakes of $H_2$ at 573 K[26]. As previously suggested for MOFs, metal doping enhances the storage efficiency of nanotubes tremendously. Carbon allotropes in all dimensions (0D-fullerenes, 1D-CNTs, 2D-graphene, graphyne etc) has been reported to enhanced their gravimetric weight percentage with metal decoration. Chandrakumar et al., in 2007, have reported that alkali metals such as Li, Na doped on fullerene can remarkably boost the weight percentage of hydrogen intake up to ~9.5 wt% [27]. In Yttrium decorated $C_{24}$, $6H_2$ molecules can be adsorbed with average 477 K desorption and $H_2$ gravimetric wt% of 8.84 [28]. Yttrium decorated $C_{60}$ can adsorb $4H_2$ molecules resulting in $H_2$ gravimetric wt% of 6.30 [29]. Metal decorated various fullerenes $C_{20}$, $C_{24}$, and $C_{60}$ have been computationally explored; for instance, Li and Na decorated $C_{20}$, Li and Na decorated $C_{20}$ [30] alkali and alkaline decorated $C_{24}$ [31] Pd and Co decorated $C_{24}$ [32] Akali metal decorated $C_{60}$ [33] $C_{60}$ impregnated in Metal-organic frameworks [34]. Like fullerene and CNT, two-dimensional nanomaterials also exhibit a high $H_2$ gravimetric weight percentage and have the advantage of adsorbing hydrogen on both sides. Yttrium doped single-wall carbon nanotubes (SWCNT) and graphyne tubes are reported to uptake $H_2$ of 6.1 wt% and 5.73 wt% [35,36]. Other metals such as Li, Be, Na, and Ru decorated CNTs have been reported [37,38].

Graphene is one of the most promising two-dimensional materials because of its unique combination of features. It has been explored for hydrogen storage with and without modification (doping) of its surface. Ataca et al., 2008 have found that four hydrogen molecules per Li atom can be stored on Li+graphene, resulting in 12.8 wt% [39]. However, these calculations have been performed using ultrasoft pseudopotential, which usually overbinds hydrogen molecules, i.e., it estimates more adsorbed hydrogen molecules per metal atom. Ca decorated zig-zag graphene nanoribbon is measured gravimetric capacity of 5 wt% [40]. Also, Zr decorated graphene with 11 wt% was reported on an average desorption temperature of 433 K [41]. Yttrium decorated graphene can adsorb at most $6H_2$ molecules resulting in $H_2$ uptake of 7.58 wt% [42]. Mg, Ca, Cu, and Pd decorated on graphene sheet [43,44] and metal dispersed on porous graphene [45] have been reported. Moving towards high porous carbon nanomaterials, many graphenes like structures, e.g., graphyne nanotube, graphyne (GY), graphdiyne (GDY), boron-doped graphdiyne (BGDY), etc., have been discovered and explored for hydrogen storage capability [46,47]. Yttrium decorated graphyne has been reported to adsorb 9H2 molecules with the help of acetylene bond contribution [48]. Li, Ca, Sc graphyne nanotube and graphyne sheet have been reported [49–52]. Recently, an advanced graphene family member with penta, hexa, and hepta rings, namely ψ-graphene, is hypothesized [53]. It was computationally demonstrated that ψ-graphene decorated with Ti-atom is one of the best hydrogen storing materials, with a gravimetric weight percentage of 13.1 [54]. Thus, it has been theoretically demonstrated that graphene and its family members possess a high capability for hydrogen storage.

As far as experiments are concerned, experimental values do not observe the same as theoretically predicted values in most cases. There are two-fold challenges for these issues: 1-In most simulation works, stability and clustering have not been analyzed, which is one of the main features of our work, 2- While performing experiments, purity of nanomaterial, strong binding between decorated atom and nanomaterial should ensure. Apart from this, oxidation of nanomaterial and random decoration of metals should be avoided. We believe that considering these issues would help experimentalists to realize much better results. Samantaray et al. have reported the hydrogen storage weight percentage of 4.2 at 298 K, under 30 bar, for Pd-Co alloy decorated on nitrogen/boron doped graphene oxide[55]. Al/Ni/graphene composites demonstrated maximum adsorbed $H_2$ weight percentage up to ~5.7 at 473 K, with desorption efficiency of 96-97% [56]. Mg and its composites, e.g., Mg/MgH$_2$ decorated on Ni doped graphene, show a high reversible hydrogen capacity of more than 6.5 wt% [57].

Holey graphyne is a newly synthesized two-dimensional carbon allotrope obtained by using Castro-Stephens coupling reaction from 1,3,5-tribromo-2,4,6-triethynylbenzene[58]. It has uniform pores of six, eight, and twenty-four vertex rings surrounded by zero, two, and six acetylene linkages (sp$^1$-

$sp^1$ bonds). Two carbon atoms of one benzene ring attach with two carbon atoms of an adjacent octagon. Since it has a band gap of ~1 eV, several pores, and an $sp^1$-$sp^1$ bridge, it is expected to exhibit electronic, optoelectronic, energy, hydrogen storage, catalytic, and many other properties. Geo et al. have reported the Li (alkali metal) decorated holey graphyne as a potential hydrogen storage material [59].

However, so far there are no reports of transition metal yttrium (Y) doped holey graphyne sheet being considered as a potential hydrogen storage material. Therefore, in the present work, the hydrogen storage capability Y-modified holey graphyne is explored using first-principles, density-functional theory. Y-atom is selected because it has only one d-electron, which contributes to better hydrogen storage [60]. We performed vibrational calculations to ensure the stability of HGY sheet and Y decorated HGY. Then we demonstrate that the binding of Y atom on HGY is strong enough to make the HGY+Y system stable, and each Y can store up to seven hydrogen molecules. Further, to check the clustering and stability of Y doped HGY, diffusion energy barrier calculations and ab-initio molecular dynamics simulations have been performed. Our results suggest that Y-doped holey graphyne can be a valuable medium for onboard hydrogen storage devices.

## 2. Computational Details:

All the calculations have been performed employing the first-principles density functional theory (DFT) formalism as implemented in the VASP package [61,62], which utilizes the projector-augmented wave (PAW) method for pseudopotentials. Considering that generalized gradient approximation (GGA) results are more proximate to experimental values than the ones obtained using the local-density approximation[63,64], we simulated all the steps using GGA exchange functional [64], with the plane-wave cutoff energy and the Monkhorst-pack grid k-point cutoff of 500 eV, and 5x5x1, respectively. However, for AIMD we consider only single gamma point sampling (1x1x1) of the Brillouin zone [65]. To avoid interactions between different layers, a vacuum of 15 Å was used. Self-consistent field convergence was chosen to be $10^{-6}$ eV for consecutive energy steps. For geometry optimization, the ionic relaxation was performed until the Hellman-Feynman force felt by each atom was smaller than $10^{-2}$ eV/Å. For phonon calculation, we increased energy cutoff and force convergence criteria to $10^{-8}$ eV and $10^{-3}$ eV/A, respectively, and used phonopy to plot band spectrum [66]. In our calculations, 2x2 supercell of holey graphyne was selected, containing a total of 96 atoms. The valence configurations of C, Y, and H for pseudopotentials in PAW are taken as $2s^22p^2$, $4s^24p^64d^15s^2$, and $1s^1$, respectively. To incorporate van der Waals interaction which mainly contributes to physisorption, we repeated all the calculations with Grimme'correction, DFT+D2 [67]. Furthermore, molecular dynamics simulations [68,69] were performed in two steps to ensure structural stability at higher temperatures: first, using the standard molecular dynamics the temperature was raised from 0 K to 300 K and the highest desorption temperature within a fixed number of particles, constant energy, constant volume (NVE) ensemble. Second, with a fixed number of particles, constant volume, and constant temperature (NVT) ensemble employing Nose-Hoover thermostat [70], we observe the structures for 5 ps, with time-steps of 1 fs, both at room temperature (300 K) as well as the highest desorption temperature.

## 3. Results and Discussion

**3.1 Optimized holey graphyne:** Geometrically relaxed holey graphyne (HGY) structure has been plotted in Fig. 1(a). After DFT optimization, we performed phonon calculation of monolayer of HGY to check its thermal stability and plotted in Fig. 2(a), matching nicely with reported literature [59]. All Positive frequencies indicate the stability of HGY sheet. After ensuring the HGY stability, we compare it with similar structures and analyze it before beginning hydrogen storage simulations. This structure can be seen as an analog of holey graphene. The differences between holey graphene and holey graphyne are: holey graphene has all $sp^2$ hybridized carbon atoms while holey graphyne has two kinds of carbon atoms, $sp^1$ and $sp^2$ hybridized; holey graphene has two kinds of pores, i.e., hexagonal and 24-vertex holes, while holey graphyne has three kinds of pores of six, eight, and 24 vertex pores. Also, we can compare the structural difference between graphyne and holey graphyne. Graphyne has the biggest hole of 12 vertexes surrounded by three acetylenic bonds, while holey graphyne has 24 vertex holes surrounded by six acetylenic linkages. In graphyne, two adjacent benzene rings are connected with an acetylenic linkage, while in holey graphyne, they are connected with an eight-vertex ring. HGY has four different kinds of bonds, B1, B2, B3, and B4, shown in Fig. 1(a), resulting from $sp^2$-$sp^2$ (between a hexagon and an octagon pore), $sp^2$-$sp^2$ (between a hexagon and a 24-vertex pore), $sp^1$-$sp^2$, and $sp^1$-$sp^1$ hybridization. Our optimized parameters values match the previous literature as given in Table 1. Also, holey graphene and holey graphyne have two kinds of carbon atoms, $sp^1$ and $sp^2$ hybridized.

**3.2 Decoration of Y atom on HGY:**

$H_2$ doped on pristine HGY is optimized at two different positions (top of hexagon and octagon). But their binding energies (~0.11eV) are out of DoE criteria ranges (0.2-0.7eV), showing extremely weak binding of HGY to $H_2$ as reported in previous littreture [59]. Hence $H_2$ adsorption at room temperature seems impossible and pristine HGY is unsuitable for hydrogen storage. We have plotted the relaxed structure of HGY+$H_2$ in Fig. S1. However, it has also been shown that several nanomaterials show good hydrogen storage efficiency after decorating with metal atoms. To improve the hydrogen storage capacity of HGY, we decorate it with yttrium atoms at holes sites H1, H2, H3, and C, shown in Fig 1(b). Symbolically, we represent them as HGY+Y(position), i.e.,

HGY+Y(H1), HGY+Y(H2), HGY+Y(H3), and HGY+Y(C) denote Y-atom decorated on HGY at H1, H2, H3, and C positions, respectively. On relaxing, we get different binding energies for different sites, as presented in Table 2. After geometry optimization, Y atoms stay at their initial sites except for the H1-site, from where it moves to the H1' site, near one of the acetylenic linkages. The red point in Fig. 1(b) shows the initial position of Y at H1, and the arrow shows where it ended up after relaxation. The adsorption energy for a given configuration is calculated from the following formula:

$$E_b(Y) = E_{HGY+Y} - E_{HGY} - E_Y \qquad (1)$$

where $E_{HGY+Y}$, $E_{HGY}$, and $E_Y$ are DFT calculated ground state energies of HGY+Y, HGY, and Y, respectively.

From Table 2, it is clear that HGY+Y(H1) is the most stable (-4.1 eV), and HGY+Y(H3) is approximately as stable (-4.0 eV) as HGY+Y(H1), while HGY+Y(H2) and HGY+Y(C) are relatively less stable (-3.1 eV and -1.1 eV ). The adsorption energies of H3 and H1 are more than Y-decorated $C_{24}$ (-3.4 eV) [28] , $C_{60}$ (-2.9 eV) [29], carbon-nanotube (-2.2 eV) [35] , graphene sheet (-2.06) [42], graphyne sheet (-3.15 eV)[48] expect graphyne nanotube (-4.99 eV) [36]. We have compared adsorption energy with our system in Table S2 (column 3). From this, we conclude that Y decorated HGY is more stable than most similar systems. The negative sign of adsorption energies of these systems indicates that these processes are exothermic. The strong binding of the former two (H1 and H3) is because the doped Y atom is near an acetylenic linkage. The strength of binding of a transition metal increases with the ratio of $sp^1$ and $sp^2$ bonds in graphyne system [71]. So on Y-atom doped at H1 and H3 would feel more influence of $sp^1$ bonds than $sp^2$, which contribute to the strong binding of Y-atom at these two positions. Since H3 is symmetric, surrounded by two acetylenic bonds, and nearly as stable as H1' case, we choose to check the hydrogen storage efficiency of HGY in which yttrium is decorated on H3 sites i.e., at the center of the octagon. Before analyzing the hydrogen storage properties of HGY+Y on H3 sites, we checked its stability by performing the phonon dispersion spectrum calculation and found all the frequencies are positive Fig. 2(b). Also, in comparing phonon dispersion of HGY and HGY+Y in Fig. 2, we found that after doping, Y-atom frequency-shifted lower because of heavy atom doping [72]. The stability of other configurations is supported by presenting their formation energy in table S1. The total density of states of HGY+Y(H3) and HGY plotted in Fig. 3 show that after doping of Y-atom on H3 position, the system transforms from a nonmagnetic semiconductor with the band gap of 0.52 eV to magnetic metal. We note that our GGA-calculated value of the band gap of 0.52 eV for the pristine HGY is in good agreement with the value 0.50 eV reported in the literature [58].

**3.3 Adsorption of hydrogen molecules on HGY+Y:**

After confirming the binding of Y to HGY, we started the study of hydrogen adsorption by putting an H$_2$ molecule at the height of 2.4 Å, measured from the Y atom of the HGY+Y system. Upon relaxation, it optimizes at a distance of 2.69 Å above the Y atom, while the H-H bond of hydrogen molecules lengthened from 0.74 Å to 0.756 Å, and the binding energy is -0.34 eV. The binding energy for the first adsorbed hydrogen molecule is calculated by using the formula:

$$BE_{H_2} = E_{HGY+Y+H_2} - E_{HGY+Y} - E_{H_2} \qquad (2)$$

where E$_{HGY+Y+H2}$, E$_{HGY+Y}$, E$_{H2}$, are the ground state energies of HGY+Y+1H$_2$, i.e., one hydrogen molecule doped on HGY+Y system, HGY+Y, and the isolated hydrogen molecule, respectively.

Next, we put two more hydrogen molecules at 2.4 Å away from Y-atom on the previous system (HGY+Y+1H$_2$), and upon relaxation, we found that the last two hydrogens are at 2.61 Å, with the H-H bond length increased from 0.74 Å to 0.76 Å, and the average adsorption energy of -0.35 eV. The average adsorption of second and third adsorbed hydrogen molecules are calculated by using the following equation [54]:

$$BE_{H_2} = \frac{1}{n}\left[E_{HGY+Y+(m+n)H_2} - E_{HGY+Y+mH_2} - nE_{H_2}\right] \qquad (3)$$

where BE$_{H2}$ is the average binding energy of H$_2$ molecules; E$_{HGY+Y+(m+n)H2}$, E$_{HGY+Y+mH2}$, E$_{H2}$ are the energy of (m+n) doped hydrogen molecules on HGY+Y, m doped hydrogen molecules on HGY+Y and isolated hydrogen molecule, respectively. We continued this process by adding two hydrogen molecules to the previous system until the binding energy of added hydrogen molecules was out of DoE criteria (0.2-0.7 eV). The average adsorption energies of 4th, 5th, and 6th, 7th hydrogen molecules are -0.42 eV, and -0.22 eV, respectively. Also, the H-H bond lengths for the 4th, 5th and 6th, 7th hydrogen molecules are 0.772 Å and 0.75 Å, respectively. All the hydrogen molecules adsorbed on the system's unit cell are plotted in Fig. 4. Interestingly, one can think of 2H$_2$, 4H$_2$, and 6H$_2$ doping and their binding energy. We introduced 1H$_2$ molecules on HGY+Y+1H$_2$, making it HGY+Y+2H$_2$, and then optimized it. Similarly, we introduced 1H$_2$ molecules on HGY+Y+3H$_2$, and HGY+Y+5H$_2$ giving us HGY+Y+4H$_2$ and HGY+Y+6H$_2$ systems and optimized them. The unit cells of 2H$_2$, 4H$_2$, 6H$_2$ absorbed on HGY+Y are plotted in Fig. S2, and their corresponding adsorption energies are given in Table 3. Thus, our simulations show that a maximum of seven hydrogen molecules can be adsorbed to the HGY+Y system. The average desorption energy per H$_2$ for our system turns out to be -0.34 eV. We have compared the average desorption energy per H$_2$ of Y-decorated other similar nanostructures with our system in Table S2 (last column). The desorption temperature corresponding to the average binding energy can be calculated using van't Hoff equations [73,74]:

$$T_d = \frac{E_b}{k_B}\left(\frac{\Delta S}{R} - \ln P\right)^{-1} \quad (4)$$

where $E_b$, $k_B$, $\Delta S$, R, and P are the average binding energy of $H_2$, Boltzmann constant, change in entropy, gas constant, and the pressure of the system, respectively. Using the above formula, the desorption temperature comes out to be 438.6 K.

**3.4 Computation of Gravimetric Hydrogen Uptake:**

The metal loading pattern decides the gravimetric wt% for the absorbed $H_2$. Metals should not be loaded very close to each other to prevent metal-metal clustering. We can consider the doping of Y atoms at different positions on the HGY system, but the octagon positions (H3) are found to be favorable from the above calculations. So, we can put a Y atom at the top of every octagon. This will result in 3 Y atoms per unit cell and 21 hydrogen molecules per unit cell on one side while the same number of Y atoms and hydrogen molecules can be attached to the other side, i.e., $24C+6Y+42H_2$ per unit cell. Metal loading patterns without and with adsorbed hydrogen molecules are plotted in Fig. 5 and Fig. S3, respectively. In Fig. S3, hydrogen adsorbed on Y decorated HGY sheet is optimized, showing that the maximum distance between Sc-$H_2$ molecules is ~3.8 Å which is less than the previous reported effective distance of 4 Å for considering $H_2$ molecules to compute wt%. Hence all the $7H_2$ molecules per Y atom are considered for gravimetric weight percentage. Consequently, gravimetric density turns out to be 7.08 wt% for the one-sided adsorption and 9.34 wt% for adsorption on both sides. The desorption temperatures and gravimetric densities of various similar materials are also compared in Table 4.

Since we predict that a Y atom can be doped on each eight-vertex pore, we need to make sure that Y atoms are not attracted to each other such that they can form clusters which may reduce the hydrogen adsorption capacity of HGY+Y system. The adsorption energy (-4.0 eV) of Y+HGY is greater than that of a similar system Y+graphyne (3.15 eV) in magnitude, implying the probability of clustering in Y+HGY is lower than that of Y+graphyne. However, the cohesive energy of bulk Y is more than the adsorption energy of Y to HGY. Hence, to prevent clustering, there should be some barrier between H3 position and the other positions. We considered two paths, P1 (from H3 position to C) and P2 (from H3 position to H2 position), to check the availability of energy barriers. We performed single point total energy calculations of different configurations along the path P1 and P2 and plotted them in Fig. 6. From Fig. 6, it is clear that there are considerable barriers of 2.95 eV and 13.9 eV, respectively, along the two paths. These barriers must be higher than the thermal energy of the highest desorption temperature; otherwise, clustering of Y atoms can occur in the process of desorption of hydrogen. The thermal energy at the desorption can be calculated using the expression $1.5k_BT$, where T is the highest desorption temperature. The highest desorption temperature for our system is 541 K, which corresponds to 0.07 eV, much less than the diffusion energy barriers for our system. Hence, due to these high barriers, the diffusion of a Y atom from the H3 position to other positions is improbable.

**3.5 Structural Stability from Ab-initio MD simulations:**

Solid-state hydrogen storage devices are targeted to be used at room temperature. However, our DFT calculations have been performed at absolute zero temperature. So we need to confirm the stability of our system at room temperature and the highest desorption temperature. We use the first-principles approach to molecular dynamics, known as ab-initio molecular dynamics (AIMD), to prove the stability of HGY+Y(H3) at higher temperatures. We performed these calculations in two steps: first, we increased the temperature of our system from 0 K to 300 K by using the standard molecular dynamics in the NVE ensemble. Next, in the NVT ensemble at 300 K, employing a Nose-Hoover thermostat, we performed molecular dynamics for 4.5 ps with the equal time steps of 1 fs each. After the calculation, the structure of our system HGY+Y remained the same except for a slight dislocation of Y. The structure and variation of the bond length between Y and its nearest carbon atom are plotted in Fig. 7. It shows a negligible dislocation of Y atom, and small fluctuations (~7%) in the bond length (Y-C). AIMD calculations are repeated at 400 K and 555 K (highest desorption temperature) and found that the structures remain stable even at 555 K. We plotted the variation of kinetic energy and total energy in AIMD process in Fig. 8. The variation of force components ($f_x$, $f_y$, $f_z$), the average of each component, and the total average forces acting on an atom during the AIMD process are plotted in Figs. 9 (a-d) respectively. The average of force components at a time t is computed by averaging force component before that time t, i.e., on the interval [0,t] and total average force are vector sum of average component forces. From Figs. 9(a-c) and its inset, we can see that $f_x$, $f_y$, and $f_z$ fluctuate around zero, and the average of each component is close to zero. Hence average total force on an atom is zero Fig. 9(d). The variation of forces for configurations HGY+Y(C), HGY+Y(H1), and HGY+Y(H2) are plotted in Fig. S4. The fluctuations of kinetic energy, total energy, and average force are negligible. Hence we conclude that our system is thermodynamically stable. Again, two Y atoms doped on both sides of HGY are also found to be stable at the highest desorption temperature, as shown in Fig. 10. However, synthesizing metal doped HGY may be a little tricky for the experimentalist as sometimes metal becomes metal-oxide, which may reduce the $H_2$ wt%. Metals should be decorated with low concentration so that metals will be sufficiently away from each other (separation around 4-5 Å) to avoid metal-metal clustering. Oxygen interference may be another challenge as sometimes $O_2$ gets adsorbed on the metal, introducing competition between $H_2$ adsorption and $O_2$ adsorption.

**3.6 Nature of Interaction of Y atom with HGY:**

In the upper panel of Fig. 11, we have plotted the partial density of states (PDOS) of the 2p orbital of the carbon atom nearest to the Y atom, before and after Y was attached to HGY. As is obvious from Figs. 11 (a) and (b), Y+HGY transformed to a magnetic metal when decorating Y on holey graphyne. Also, there is an enhancement of states near Fermi-level on doping Y. Similarly, in the lower panel of Fig. 11, we plotted PDOS of the 4d orbital of isolated Y, and Y attached to HGY. On comparing Figs. 11 (c) and (d), it is clear that the Y atom has a magnetic character before and after doping. The decrement of 4d states near fermi-energy implies that the Y atom has lost some charge which is transferred to the HGY sheet after binding. In order to confirm this, we performed Bader charge analysis for quantitative charge transfer and found that 1.67e charge has indeed been transferred from the Y atom to the HGY sheet. To see precisely which carbon atoms in holey graphyne have gained charge from the Y atom, we plotted the charge density difference of HGY+Y and HGY, i.e., $\rho(Y+HGY)-\rho(HGY)$ in Fig. 12. The red color around the Y atom shows the charge depletion region, while blue and light green show charge gain regions. On the color scale, blue represents more charge gain than green color. Hence from Fig. 12, we conclude that most of the charges have been transferred from Y atom to its nearest four C atoms in holey graphyne. Finally, we conclude that the interaction between Y atom and the HGY sheet is because of charge transfer, and, thus, the binding between the two is ionic.

**3.7 Interaction of $H_2$ with HGY+Y:** To explore the interaction between adsorbed hydrogen molecules and HGY+Y system, we plotted the PDOS of 1s H-atom of HGY+Y+1$H_2$ and the same of an isolated hydrogen molecule in the upper panel of Fig. 13. On comparing Figs. 13 (a) and (b), we can notice the density of states gets enhanced for the H 1s orbital in HGY+Y+1$H_2$ as indicated in the upper panel of Fig. 13. Hence, after the adsorption of $H_2$ molecules on HGY+Y, $H_2$ gains some charges. Similarly, we plotted PDOS of 4d-orbital of Y in HGY+Y+1$H_2$ and the same of HGY+Y in the lower panel of Fig. 13. On comparing Figs. 13 (c) and (d), decrement of 4d states can be noted. Hence, after adsorption of $H_2$ molecules on HGY+Y, Y atom loses some more charge, and the $H_2$ molecule gains that charge. Quantitatively, the same is confirmed by using Bader charge analysis which shows that each hydrogen molecule gains 0.007e charge. But on attaching more and more hydrogen on HGY+Y, the $H_2$ molecules gain smaller charges, resulting in weaker and weaker bonding between HGY+Y and hydrogen molecules. Hence, we can store a finite number of up to seven hydrogen molecules per Y atom resulting in 7.08 and 9.34 wt% if $H_2$ molecules doping is one and both sides, respectively. The adsorption energy of $H_2$ molecules on Y+HGY lies between physisorption and chemisorption ranges, so their interaction might be of Kubas-type. In this interaction, the occupied 4d-orbital of Y donates some charges to anti-sigma ($\sigma^*$) bonds of hydrogen and sigma ($\sigma$) bonds of hydrogen back donate some charge to the unoccupied 4d-orbital of Y-atom. Again, elongation of bond length of adsorbed $H_2$ molecules (see Table 3) on HGY+Y indicates that the interaction between $H_2$ and HGY+Y is of Kubas type.

## 4. Conclusion:

In summary, by employing the first principles DFT calculations, the hydrogen storage capability of ultra porous holey graphyne is found to be 7.08 and 9.34 wt% on utilizing one side and both sides of holey graphyne, with average binding energy and desorption temperature -0.34 eV and ~438 K, respectively. The interaction between Y and HGY is due to charge transfer as analyzed from the partial density of states and Bader charge analysis. Furthermore, the interaction between $H_2$ molecule and HGY+Y system is of Kubas type. Diffusion energy barriers for metal atoms along the possible diffusion paths are calculated and found to be large enough to prevent diffusion and clustering, a property very important for the practical feasibility of this system. Also, stability at room temperature and the highest desorption temperature was checked, and a negligible dislocation of Y-atom was found in the HGY-sheet. Therefore, we strongly believe that holey graphyne sheet can be useful as an onboard hydrogen storage medium when used in conjunction with the doping by yttrium atoms.

ACKNOWLEDGEMENTS

Mukesh Singh would like to express his special thanks to CSIR for funding his Ph.D. research projects. Mukesh Singh also acknowledges the support of Spacetime High Performance Computing team and its facility at IIT Bombay.

**Figures and tables**

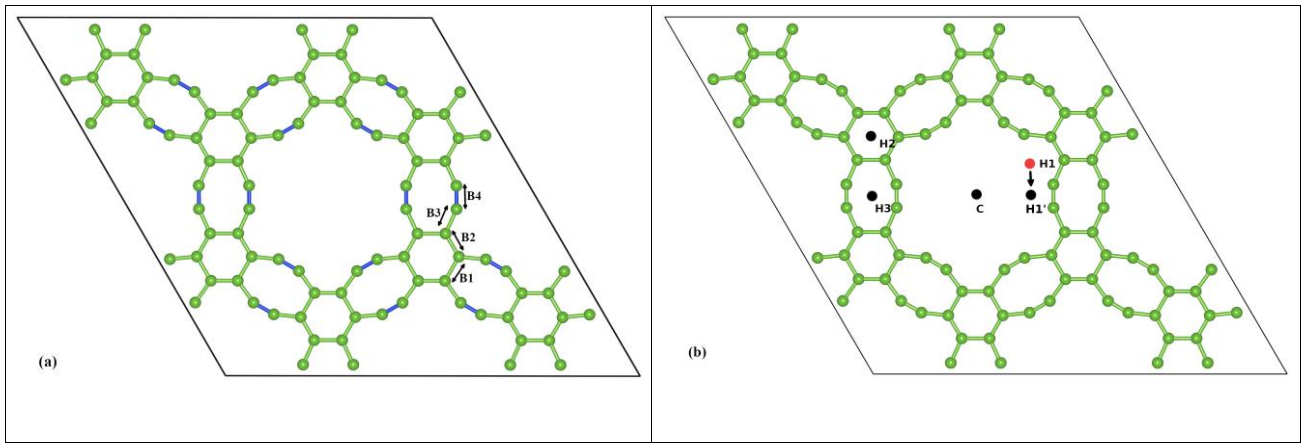

Fig. 1: (a) DFT Relaxed holey graphyne structure. Green spheres represent the carbon atoms. Green and blue rods represent $sp^2$ and $sp^1$ bonds respectively. B1, B2, B3, B4 are four different bond length with the values of 1.46, 1.40, 1.41, 1.23 Å, respectively. (b) Doping of Y atoms at different positions on the holey graphyne layer. The red and black dots show positions of Y atoms before and after the relaxation of Y+graphyne. In case of position H1, red dot and black dot are different points so after relaxation Y-atom has moved which is shown by black arrow. For H2, H3, and C positions, Y-atom remains at the same postion so that red and black dots coincide.

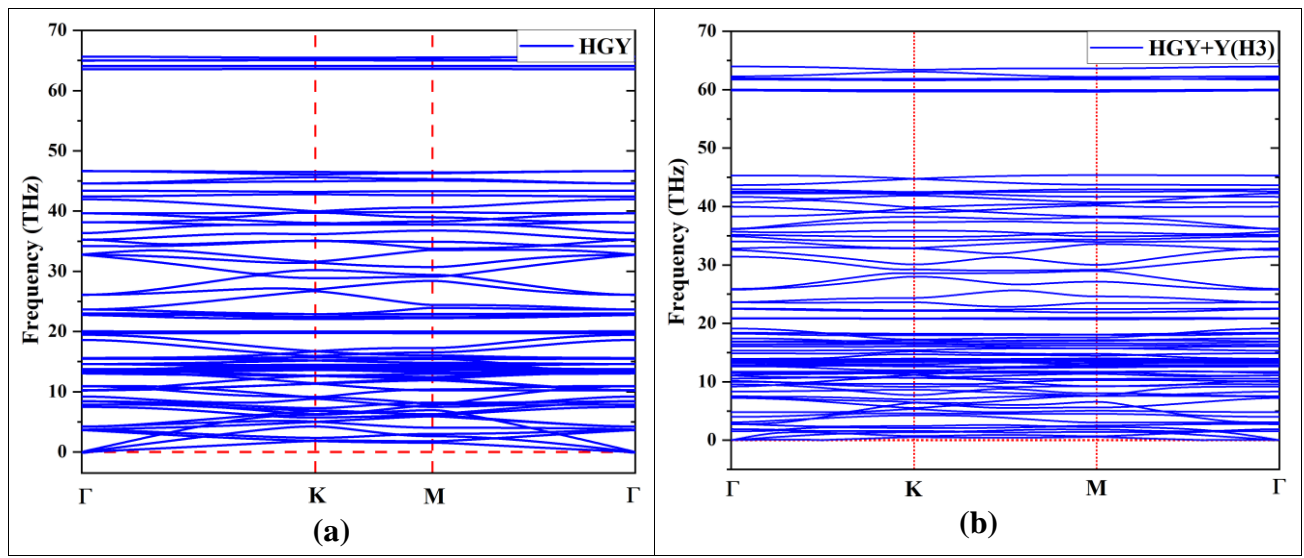

Fig. 2: Phonon (vibrational) spectrum of: (a) monolayer HGY sheet (b) monolayer HGY+Y (H3). After doping Y on HGY sheet, phonon spectrum shifted to lower frequencies, i.e., acoustic mode comes close to fermi-level, and the highest optical mode shifted to below 65 THz.

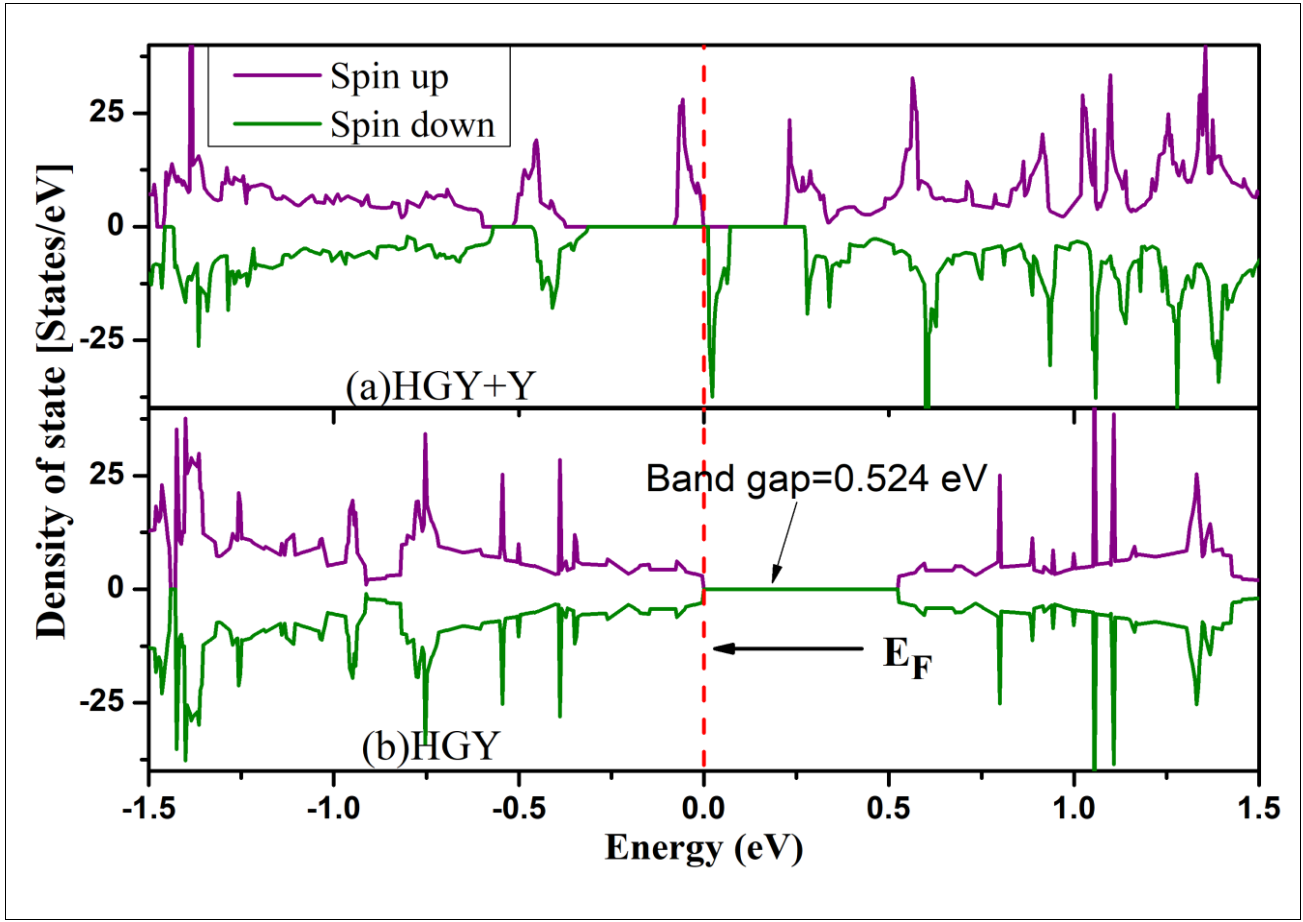

Fig. 3: Density of states before (upper panel) and after(lower panel) doping Y-atom on holey graphyne sheet. Fermi energies are equated to zero

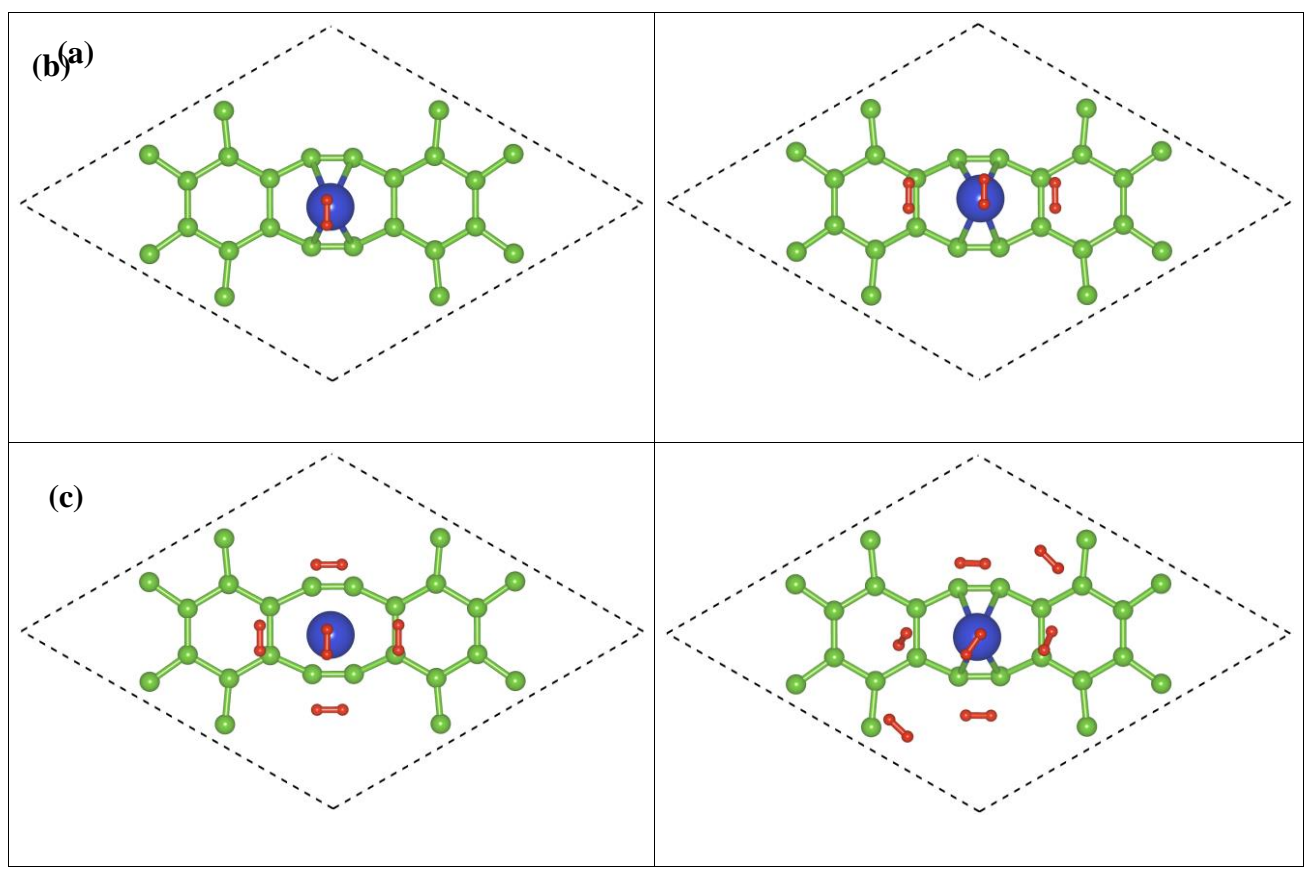

Fig. 4: Unit cell of the relaxed structure of HGY+Y molecules, with varying numbers of $H_2$ molecules adsorbed: (a) one hydrogen adsorbed, i.e., HGY+Y+1$H_2$, (b) HGY+Y+3$H_2$, (c) HGY+Y+5$H_2$, and (d) HGY+Y+7$H_2$. The green, blue and red spheres represent the carbon atoms of HGY, yttrium atom, and hydrogen atoms, respectively.

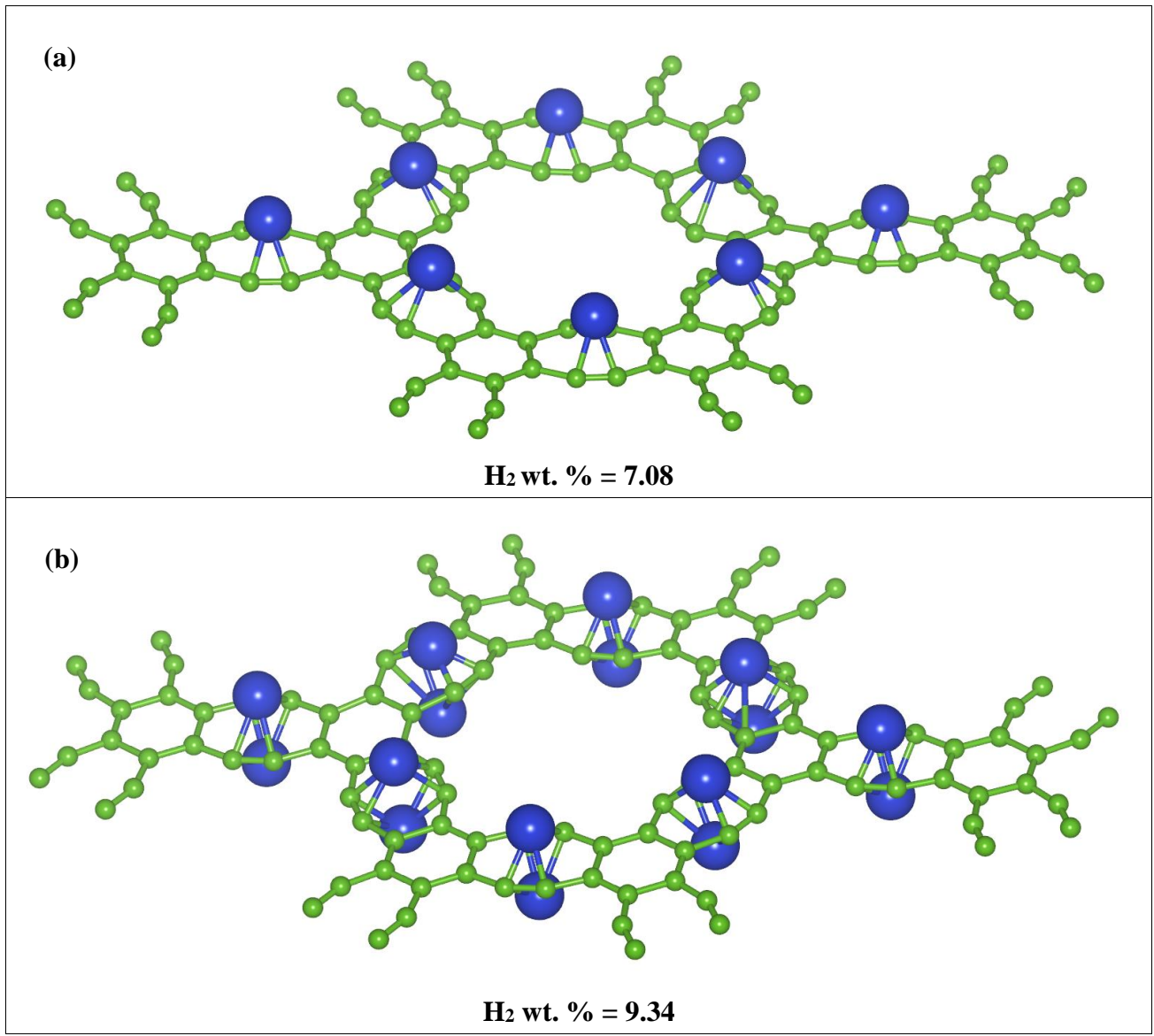

Fig. 5: Various metal (Y) loading configurations along with the corresponding gravimetric weight percentage of: (a) when Y is doped on one side (b) when Y is doped on both the sides of holey graphyne.

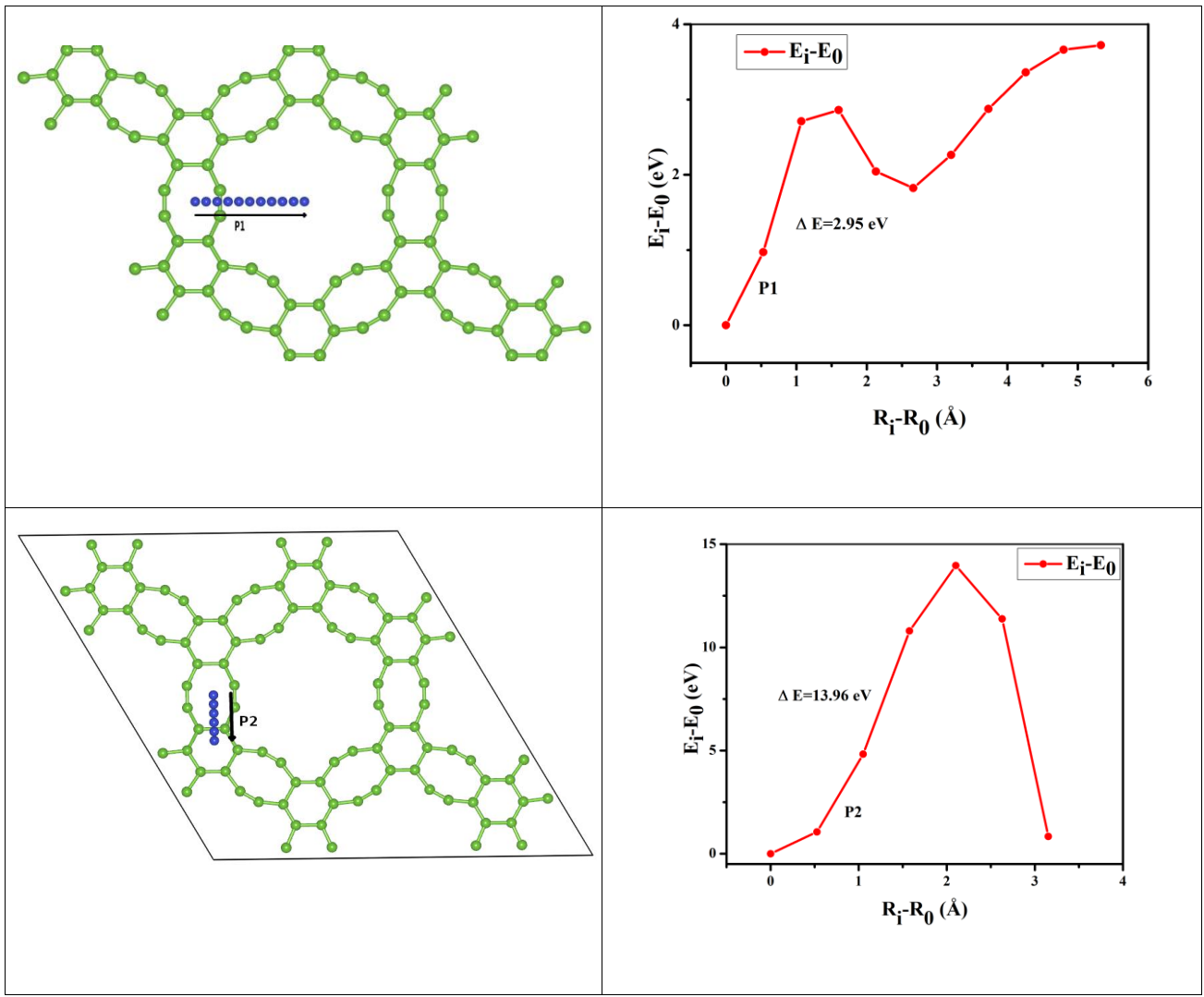

Fig. 6: Variation of energy difference ($E_i - E_0$) as a function of the displacement ($R_i - R_0$) of the yttrium atom from the center of the octagon, to the center of hexagon, and that of the supercell, along paths P1, and P2, respectively.

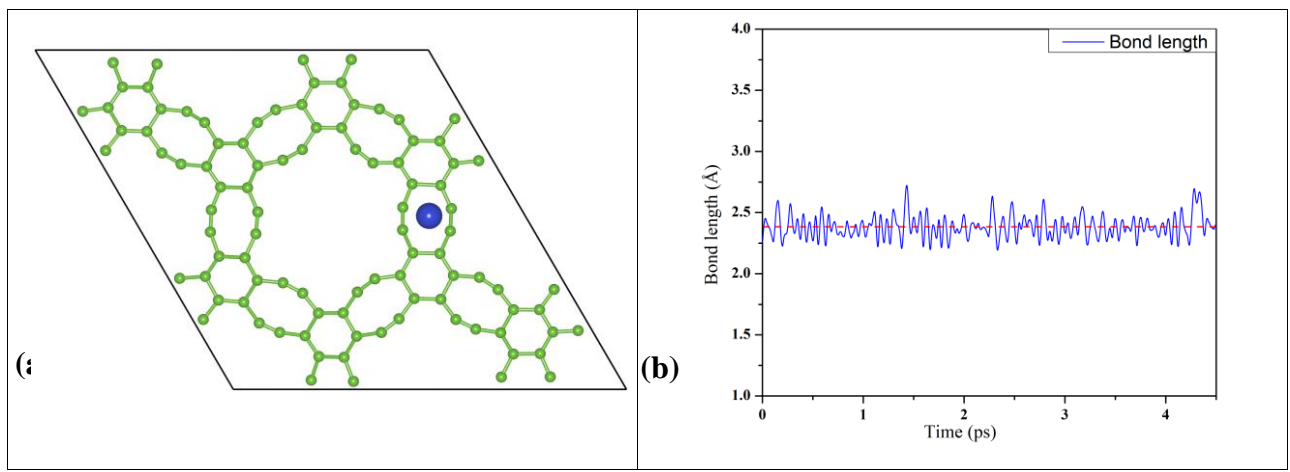

Fig. 7: (a) Molecular dynamics snap shots of HGY+Y after 4.5 ps at 555 K. The green and the blue spheres are the carbon atoms of HGY sheet, and the yttrium atom, respectively. The yttrium atom has a minute displacement form its initial position. (b) The variation in the bond length of Y atom with respect to the nearest carbon atom of the HGY sheet is ~7 %.

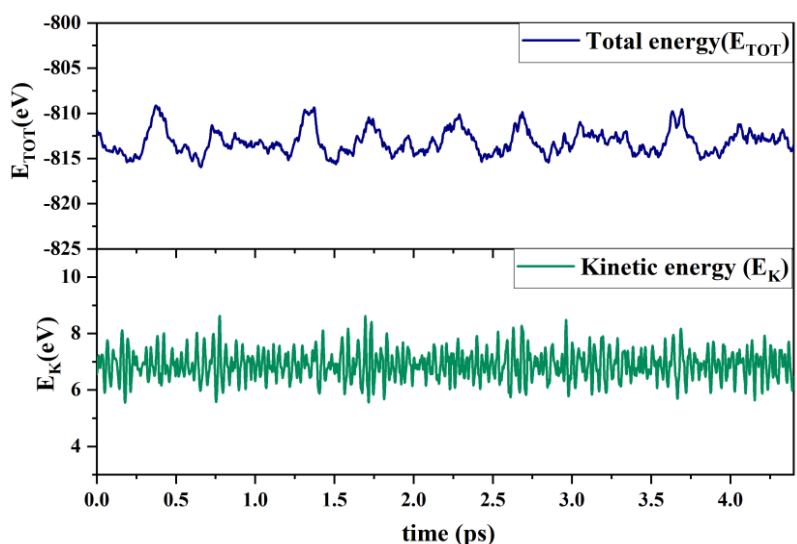

Fig. 8: At highest desorption temperature 555K, system variation of kinetic and total energy plot at each iterations in 4.5 ps with step of 1fs for HGY+Y.

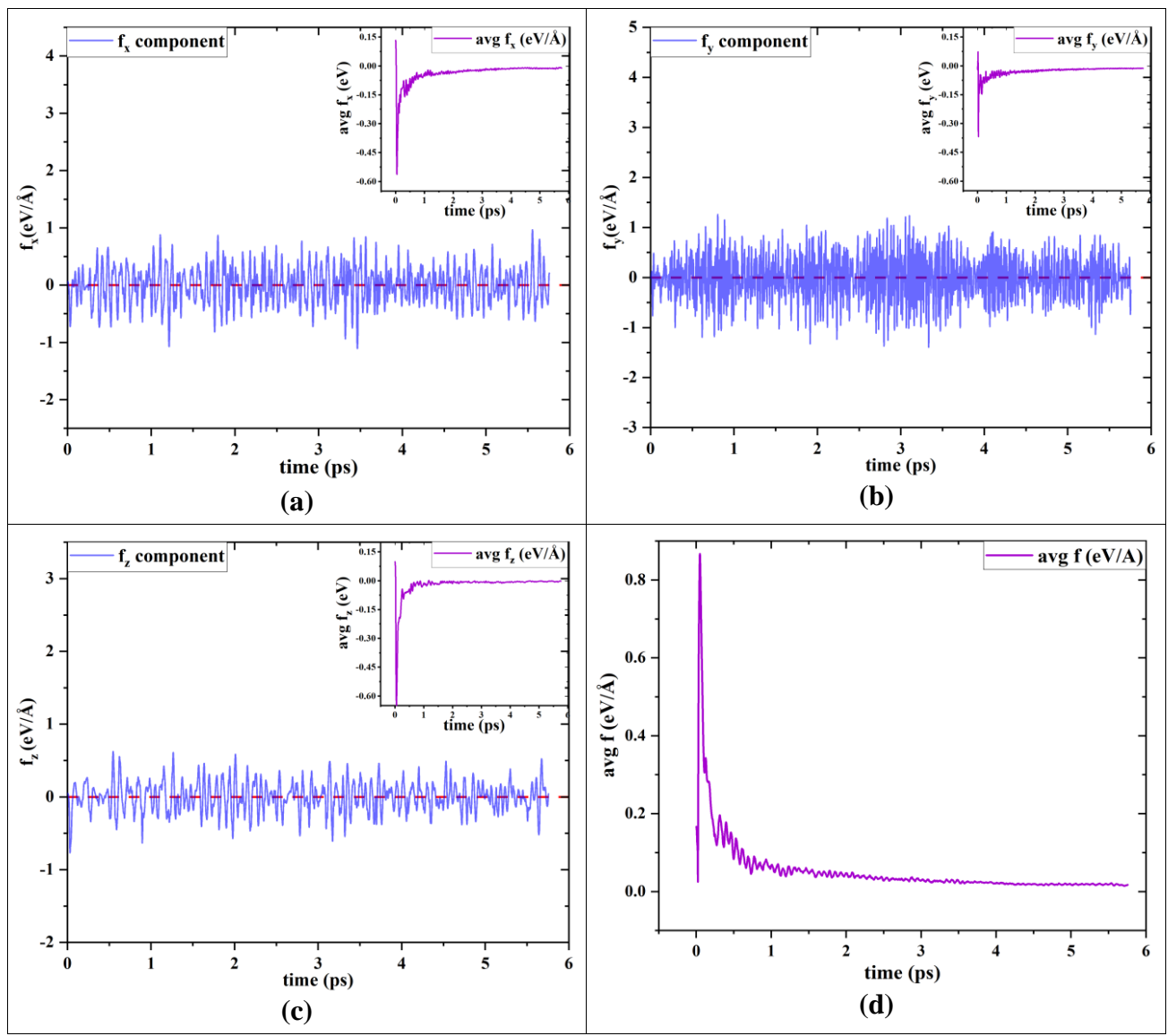

Fig. 9: During AIMD at 300K for HGY+Y(H3), fluctuation $f_x$, $f_y$, $f_z$-components of forces and Total average force plotted in **a, b, c, d**; average component forces are plotted insets.

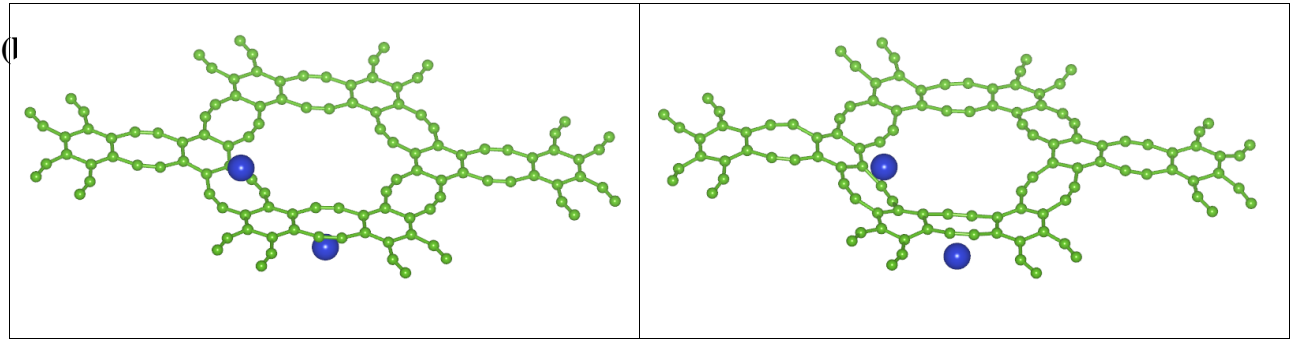

Fig. 10: (a) HGY+2Y at the start of molecular dynamics simulations, with Y atoms attached on the opposite sides of the sheet, above/below centers of the two nearest octagons, and (b) molecular dynamics snapshot of HGY+2Y after 5ps at 400 K. Negligible displacement in the location of the Y atoms is seen. The green and blue sphere are carbon atoms of the HGY sheet, and the Yttrium atom, respectively.

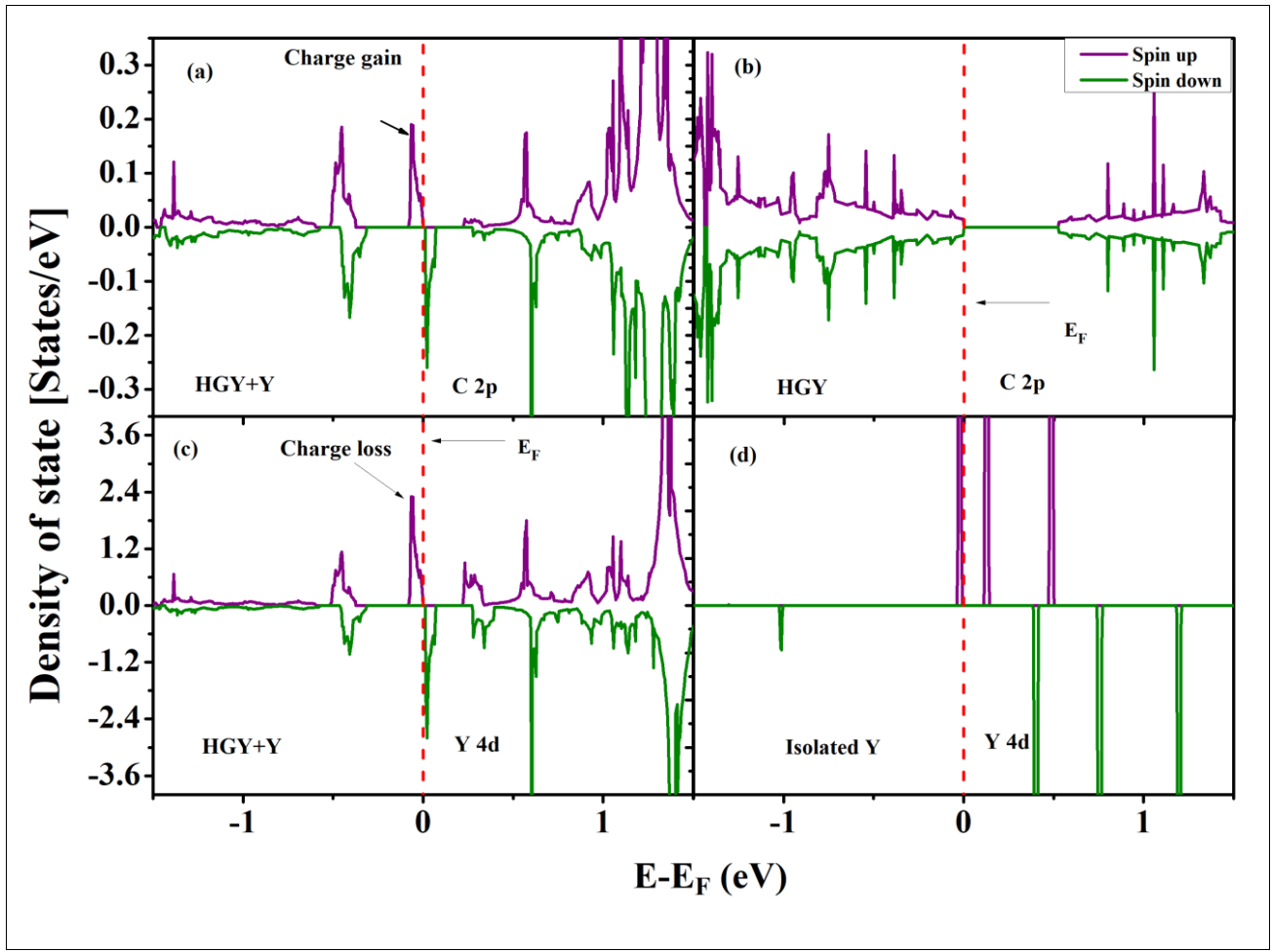

Fig. 11: Partial density of states of: (a) 2p orbital of carbon atom in HGY+Y; (b) 2p orbital of carbon atom of pristine HGY; (c) 4d orbital of Y atom of HGY+Y; (d) 4d orbital of isolated yttrium atom. On comparing (a) and (b) we note that some charge has been gained by holey graphyne sheet. While on comparing (c) and (d) it is obvious that charges are transferred from Y-atom, to the HGY sheet. Fermi energies are equated to zero.

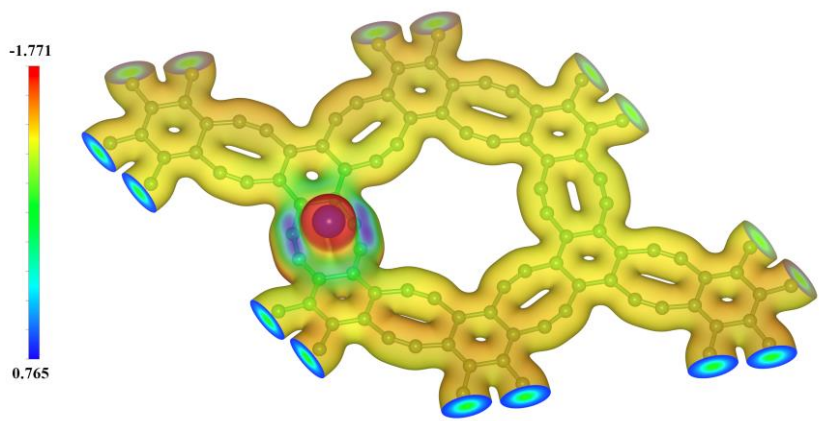

Fig. 12: Difference charge density plot of ρ(HGY+Y)-ρ(HGY) with the isovalue 2.84 x $10^{-2}$. Red and blue colors represent positive and negative charge regions implying that most of the charge is being transferred from yttrium to four carbon atoms nearest to it.

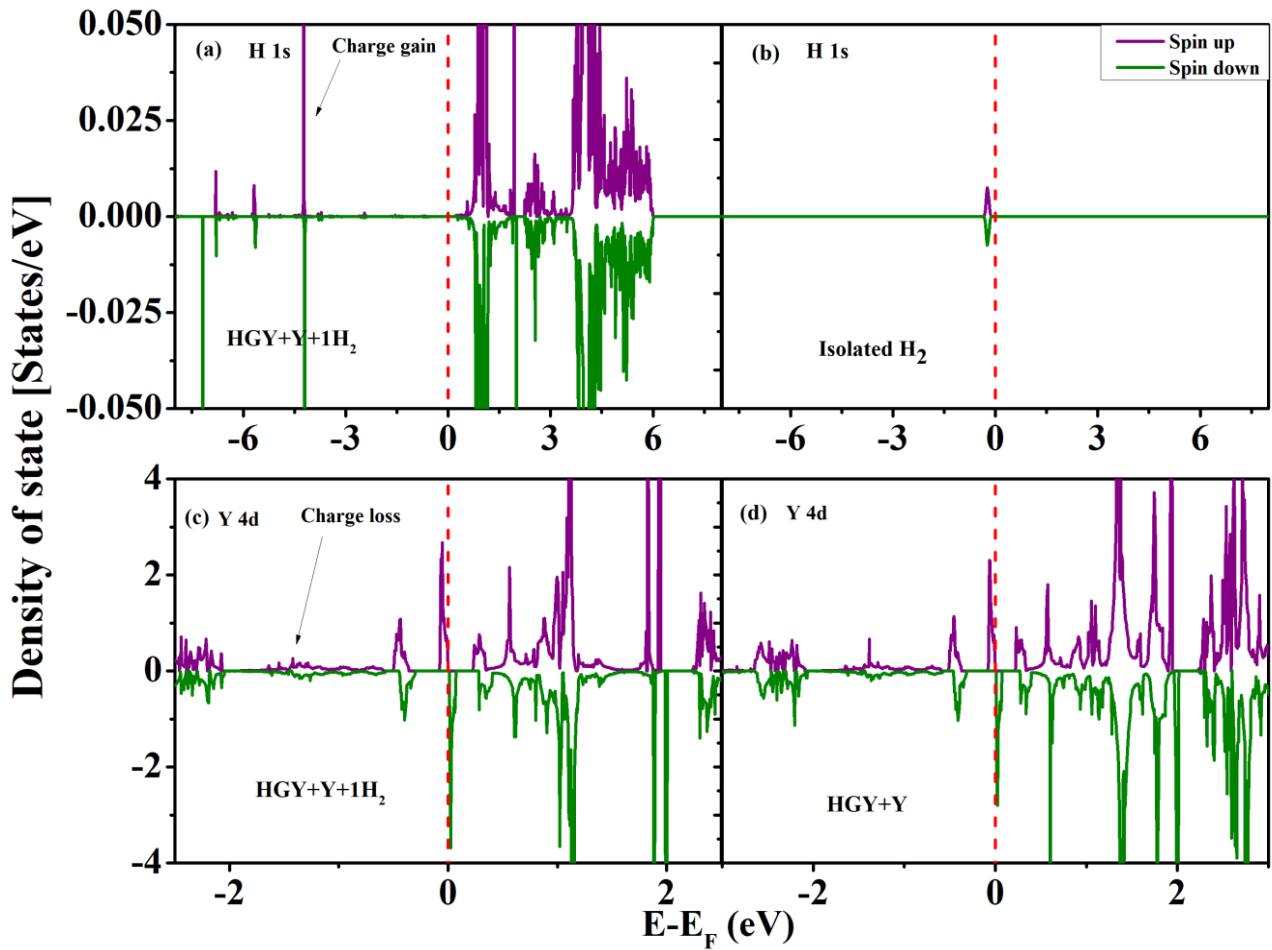

Fig. 13: Partial density of states of: (a) 1s orbital of hydrogen atom of HGY+Y+1H$_2$; (b) 1s orbital of hydrogen atom of isolated H$_2$ molecule; (c) 4d orbital of yttrium atom of HGY+Y+1H$_2$; (d) 4d orbital of yttrium atom of HGY+Y. It is obvious that some charge is transferred from Y atom to the H$_2$ molecules. Fermi energies are equated to zero.

Table 1: Structural properties and the band gap of DFT optimized holey graphyne.

| S.N | Properties | Present work | Reported result [58, 59] |
|---|---|---|---|
| 1 | B1 | 1.463 Å | 1.463 Å |
| 2 | B2 | 1.398 Å | 1.397 Å |
| 3 | B3 | 1.413 Å | 1.414 Å |
| 4 | B4 | 1.226 Å | 1.227 Å |
| 5 | ∠ABC | 125.6º | 125.8º |
| 6 | Band-gap (PBE) | 0.513 eV | 0.50 eV |

Table 2: Binding strength of Y atom attached at different positions on holey graphyne. These position are indicated in Fig 2.

| Position of Y-atom before relaxation | H1 | H2 | H3 | C |
|---|---|---|---|---|
| After relaxation | H1'(moved) | H2 | H3 | C |
| Binding energy (eV) | -4.10 | -3.10 | -4.00 | -1.10 |

Table 3: Binding energies of $nH_2$ doped systems computed using the GGA functional including the van der Waal corrections with Grimme-D2. H-H bond lengths of the $H_2$ molecules are also reported.

| S.N. | System | Corrected DFT+D2 binding energy (eV) | H-H bond legth of doped hydrogen molecules after relaxation (Å) |
|---|---|---|---|
| 1 | HGY+Y | -4.00 | -- |
| 2 | HGY+Y+1$H_2$ | -0.34 | 0.756 |
| 3 | HGY+Y+2$H_2$ | -0.37 | 0.762 |
| 4 | HGY+Y+3$H_2$ | -0.35 | 0.760 |
| 5 | HGY+Y+4$H_2$ | -0.43 | 0.785 |
| 6 | HGY+Y+5$H_2$ | -0.42 | 0.772 |
| 7 | HGY+Y+6$H_2$ | -0.25 | 0.751 |
| 8 | HGY+Y+7$H_2$ | -0.22 | 0.750 |
| Average binding per $H_2$ = -0.34 eV | | | |
| Average desorption temperature = 438.6 K | | | |

Table 4: Summary for Gravimetric hydrogen percentage and desorption temperature of hydrogen storage materials predicted by DFT and experimentally tested holey graphyne like materials.

| S.N. | System | Average desorption temperature(K) | $H_2$ wt% |
|---|---|---|---|
| 1. | Li+ graphene [39] | 271 | 12.8 |
| 2. | Ca+zigzag graphene nanoribbon [40] | 258 | 5.0 |
| 3. | Zr+graphene [41] | 433 | 11.0 |
| 4. | Li+graphyne [47] | 348 | 18.6 |
| 5. | Li+holey graphyne [59] | 283 | 12.8 |
| 6. | Y+fullerene ($C_{24}$) [28] | 477 | 8.84 |
| 7. | Y+ carbon nanotube [35] | 396 | 6.10 |
| 8. | Y+ graphyne nanotube [36] | 197 | 5.73 |
| 9. | Y+graphene [42] | 725 | 5.78 |
| 10. | Y+graphyne [48] | 400 | 10.00 |
| **11.** | **Y+holey graphyne(Present work)** | **438** | **9.34** |
| | **Experimental reports** | | |
| 12. | $Pb_3Co$-NG/$Pb_3Co$-BG [55] | | ~4.2/4.6 |
| 13. | Al/Ni/graphene [56] | | ~5.7 |
| 14. | Mg/$MgH_2$+Ni/GLM [57] | | >6.5 |